%% file: s4_jbhi.tex
\newif\ifdraft
\draftfalse

\documentclass[journal,twoside, web]{ieeecolor}
\usepackage{generic}
\usepackage{cite}
\usepackage{amsmath,amssymb,amsfonts}
\usepackage{bbm}
\usepackage{graphicx}
\usepackage{hyperref}
\usepackage{caption,subcaption}
\usepackage{multirow}
\usepackage{cleveref}
\usepackage{comment}
\usepackage{hhline}
\usepackage{todonotes}
\usepackage{graphicx}
\usepackage{caption,subcaption}
\usepackage{siunitx}
\usepackage{booktabs}
\usepackage{csvsimple}
\usepackage{datatool}
\usepackage{array}

\usepackage[normalem]{ulem} %for sout; normalem to remove issues with underlined journal names in bibstyle
\newcommand{\stkout}[1]{\ifmmode\text{\sout{\ensuremath{#1}}}\else\sout{#1}\fi}
\ifdraft
\newcommand{\added}[1]{\textcolor{blue}{#1}}
\newcommand{\deleted}[1]{\textcolor{red}{\stkout{#1}}}
\newcommand{\replaced}[2]{\textcolor{blue}{#1} \textcolor{red}{\stkout{#2}}}
\newcommand{\deletedfloat}[1]{}%{\textcolor{red}{#1}}%for tables
\newcommand{\commented}[1]{\textcolor{blue}{#1}}
\else
\newcommand{\added}[1]{#1}
\newcommand{\deleted}[1]{}
\newcommand{\replaced}[2]{#1}
\newcommand{\deletedfloat}[1]{}%for tables
\newcommand{\commented}[1]{}
\fi

\newcommand\citep[1]{\cite{#1}}
\newcommand\citet[1]{\cite{#1}}

\def\BibTeX{{\rm B\kern-.05em{\sc i\kern-.025em b}\kern-.08em
    T\kern-.1667em\lower.7exE\kern-.125emX}}
\begin{document}

\title{Towards quantitative precision for ECG analysis: Leveraging state space models, self-supervision and patient metadata}
\author{Temesgen Mehari and Nils Strodthoff
\thanks{Temesgen Mehari is with Physikalisch-Technische Bundesanstalt, Berlin, Germany and Fraunhofer Heinrich Hertz Institute, Berlin, Germany (email: temesgen.mehari@hhi.fraunhofer.de). Nils Strodthoff is with Oldenburg University, Oldenburg, Germany. (email: nils.strodthoff@uol.de). Corresponding author: NS. This project (18HLT07 MedalCare) has received funding from the EMPIR programme co-financed by the Participating States and from the European Union’s Horizon 2020 research and innovation programme.}}

\bstctlcite{BSTcontrol}

\maketitle

\begin{abstract}
    Deep learning has emerged as the preferred modeling approach for automatic ECG analysis. In this study, we investigate three elements aimed at improving the quantitative accuracy of such systems. These components consistently enhance performance beyond the existing state-of-the-art, which is predominantly based on convolutional models. Firstly, we explore more expressive architectures by exploiting structured state space models (SSMs). These models have shown promise in capturing long-term dependencies in time series data. By incorporating SSMs into our approach, we not only achieve better performance, but also gain insights into long-standing questions in the field. Specifically, for standard diagnostic tasks, we find no advantage in using higher sampling rates such as 500Hz compared to 100Hz. Similarly, extending the input size of the model beyond 3 seconds does not lead to significant improvements. Secondly, we demonstrate that self-supervised learning using contrastive predictive coding can further improve the performance of SSMs. By leveraging self-supervision, we enable the model to learn more robust and representative features, leading to improved analysis accuracy. Lastly, we depart from synthetic benchmarking scenarios and incorporate basic demographic metadata alongside the ECG signal as input. This inclusion of patient metadata departs from the conventional practice of relying solely on the signal itself. Remarkably, this addition consistently yields positive effects on predictive performance. We firmly believe that all three components should be considered when developing next-generation ECG analysis algorithms.

\end{abstract}


\input{jbhi_sections/introduction}

\input{jbhi_sections/methods}
\input{jbhi_sections/results}	

\section{Summary}
In this study, we utilized structured state space models, which are well-suited for capturing long-term dependencies in time series, to challenge the dominance of convolutional architectures in the realm of deep-learning-based ECG analysis. Our results consistently surpassed the previous state-of-the-art on large, comprehensive ECG classification datasets, both in the supervised and self-supervised settings. We leveraged the model's capability to capture long-term dependencies to shed new light on the optimal sampling frequency and input size for the model, rebuting for example common myths on necessary sampling rates. The significant performance improvements achieved by including patient metadata suggest the need to move beyond artificial benchmarking scenarios, where the model predictions are based solely on the signal, to more realistic scenarios, while maintaining a strict evaluation protocol. We are looking forward to applications of the proposed model in the context of specific clinical prediction tasks, in particular of indirect prediction tasks that are hard to accomplish for cardiologists based on the signal alone.

The code underlying our experiments is publicly available under \url{https://github.com/tmehari/ssm_ecg}.

\bibliographystyle{IEEEtran}
\bibliography{refs}

\end{document}

%% file: jbhi_sections/introduction.tex
\section{Introduction}
\noindent\textbf{Machine learning for ECG analysis}
Machine learning , in particular deep learning, has the potential to transform the entire field of healthcare. The electrocardiogram (ECG) is particularly well-suited to lead this development because of its widespread use (in the US, an ECG was ordered or provided at about 5\% of office visits \citep{NACMS2016}). While it only requires basic recording equipment, it holds enormous diagnostic potential that we only gradually start to uncover with the help of machine learning \citep{Hannun2019,Attia2019,Lima2021,Verbrugge2022}. 

\noindent\textbf{Model architectures}
On the algorithmic side, the analysis of ECGs based on raw sensor data is still largely dominated by convolutional neural networks \citep{strodthoff2020deep, hannun2019cardiologist,Attia2019,Ribeiro2020}. This default choice is slowly being challenged by the rise of transformer-based architectures or combinations of convolutional architectures with attention elements, as exemplified by the winning solutions of the two past editions of the Computing in Cardiology Challenge \citep{Natajan2020Wide,Nejedly2021}. In this work, we explore a novel algorithmic approach, \textit{Structured State Space Sequence (S4) models \citep{gu2021efficiently}}, which learn a continuous representation of time series data and are particularly suited for modeling long-term dependencies.
S4 models have demonstrated exceptional performance in modeling and analyzing long-term time series data in various domains. By leveraging the inherent sequential dependencies within ECG signals, we anticipate that the application of S4 models to the ECG domain can significantly enhance the performance of existing prediction models. By capturing complex temporal dynamics and long-term dependencies, S4 models have the potential to unlock new insights and achieve more accurate ECG analysis outcomes. We demonstrate consistent improvements in quantitative accuracy through three components: (1) the use of S4 models as internal model architecture, (2) leveraging self-supervised pretraining, and (3) the inclusion of demographic metadata, see also \Cref{fig:performance_contributions} for a visual summary of the achieved results. We discuss all three components in the following paragraphs:

\begin{figure}
	\centering
	\includegraphics[width=0.8\columnwidth]{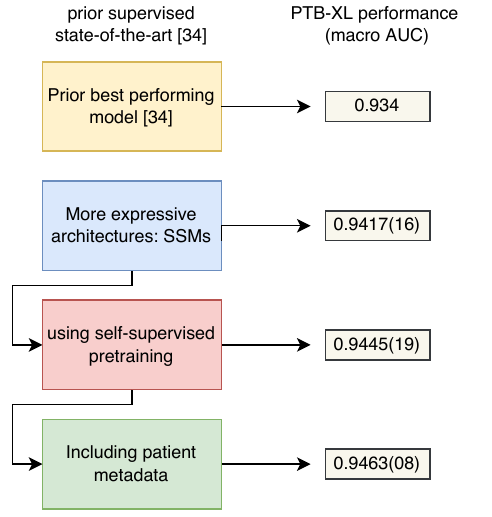}
	\caption{Visual summary: We demonstrate consistent performance improvements over the state-of-the-art through (1) the use of structured state space models (SSMs) as internal model architecture, (2) leveraging self-supervised pretraining, and (3) the inclusion of demographic metadata.}
	\label{fig:performance_contributions}
\end{figure}

\noindent\textbf{1) More expressive architectures: SSMs}
We thoroughly evaluate the model using a well-established benchmarking procedure \citep{strodthoff2020deep} on the \textit{PTB-XL} \citep{Wagner:2020PTBXL,Wagner2020:ptbxlphysionet,Goldberger2020:physionet} and \textit{Chapman} \citep{Zheng2020} datasets and show consistent improvements over the existing (convolutional as well as attention-based) state-of-the-art . Our main insight is  methodological in nature. As a consequence,  we deliberately focus  on comprehensive ECG classification tasks rather than specific clinical prediction tasks, even though we also envision that data-driven methods will eventually supplement current rule-based decision support systems in ECG devices, see also \cite{Kashou2020}.

 Furthermore, we use the model's capability to capture long-term dependencies to systematically investigate  long-standing questions in the field, i.e.\, how long-ranged are the interactions in ECG data that need to be explicitly captured  and do models actually profit from input data with a higher sampling frequency of 500~Hz as compared to 100~Hz. 

\noindent\textbf{2) Self-supervision}
Self-supervised pretraining has exhibited numerous beneficial properties in previous work, enabling models, e.g., to learn robust and data-efficient representations from unlabeled data. Building upon these findings from previous work \citep{Mehari:2021Self }, we seek to investigate the applicability of self-supervised pretraining to S4 models. By harnessing the power of self-supervised learning, we aim to further improve the performance and generalization capabilities of S4 models within the ECG domain.  \citep{Mehari:2021Self} 
While the original work used an LSTM model as the internal model architecture, we show that replacing it with causal S4 layers leads to unprecedented downstream performance on both datasets under consideration.

\noindent\textbf{3) Demographic metadata}
Previous work on automatic ECG analysis, at least in the context of deep learning, has  largely focused on identifying the most effective ways to extract useful information from the (raw) signal itself. Performing this in a comparable and reproducible fashion is a challenge by itself \cite{strodthoff2020deep}, but can only be the first step in the development of clinical decision support systems. While raw ECG data provides valuable insights, clinicians often rely on accompanying metadata to aid in the diagnostic process. Consequently, we argue that the inclusion of meta-information in the application of machine learning models is essential for comprehensive ECG analysis. This is why we propose to widen the scope of the current benchmarking activities to consider also prediction tasks that include at least basic demographic metadata that should be available to the clinician in all cases. Again, we demonstrate that the combination of S4 models, self-supervision and the inclusion of demographic metadata leads to unprecedented predictive performance.

It is worth noting, that this work builds on the material of an earlier (non-archival) workshop contribution \cite{Mehari:2022S4} and extends it by including results on self-supervised pretraining and the incorporation of patient-specific metadata.

%% file: jbhi_sections/methods.tex
\section{Related Work}
%ECG classification
\noindent\textbf{ECG classification} The field of ECG analysis is largely dominated by convolutional architectures, see \citep{Hong2020, petmezas2022state} for a recent reviews. The superiority of modern ResNet- or Inception-based convolutional architectures was confirmed in an extensive comparative study on the \textit{PTB-XL}  dataset \citep{strodthoff2020deep}. This is in line with the excellent performance of such architectures on a broad range of time series classification tasks, see \citep{IsmailFawaz2020inceptionTime}. Interestingly, this supremacy was already challenged in \citep{Mehari:2021Self}, where the convolutional baseline was outperformed by a large recurrent neural network with a fully-connected feature extractor. Therefore, it represents a natural question to ask whether architectures that are even more adapted to the necessities of time series can lead to further performance improvements.\\
%S4 related work
\noindent\textbf{Structured State Space Models for clinical time series} The motivation for the development of structured state space models (SSMs) was the wish to devise an architecture that is suited to capture long-term dependencies in very long temporal data, including medical time series as a particular example \citep{NEURIPS2021_05546b0e}. To support the applicability to the latter, the authors considered a classical vital sign prediction task on ECG and photoplethysmography time series as input \citep{NEURIPS2021_05546b0e,Gu2022hippo2} and clearly outperformed the current state-of-the-art for this tasks. These results represent a very encouraging sign for the application of these models in the broader context of medical time series. Nevertheless the prior study cannot be considered as a comprehensive ECG analysis task, which is the topic of this work.

In a different line of work, SSMs were used to model the internal state in diffusion models for time series imputation \citep{juan2022}, which lead to unprecedented imputation quality (among others) for ECG data, which provides additional hints for the potential advantages of SSMs also in a purely discriminative setting. We therefore aim to investigate the motivating claims for SSMs in the context of ECG data.

%ECG self-supervised
\noindent\textbf{Self-supervised learning for ECG data} Driven by the recent success of self-supervised learning in natural language processing \citep{devlin-etal-2019-bert}, speech \citep{baevski2020wav2vec} and most recently also computer vision \citep{chen2020simple}, there have been several studies that applied related techniques also in the field of ECG analysis \citep{Yuan2019,cheng2020subjectaware,Sarkar2020,pmlr-v139-kiyasseh21a,pmlr-v158-gopal21a, lan2022intra}. Most of these studies show a rather strong methodological focus and clearly demonstrate the advantages of self-supervised pretraining as compared to conventional supervised training. However, as many of the used models tend to be shallow, the corresponding supervised baselines often fail to reach state-of-the-art performance on comprehensive ECG classification tasks and it therefore remained unclear if and to what degree these improvements would carry over to state-of-the-art models. In \citep{Mehari:2021Self}, using an adaptation of contrastive predictive coding (CPC) \citep{oord2018representation} established for representation learning for speech data, it was shown that self-supervised pretraining can in fact lead to statistically significant performance improvements compared to the state-of-the-art based on supervised training. To this end, the authors used an adaptation of contrastive predictive coding \citep{oord2018representation}, which has been established for representation learning in the context of speech. These performance improvements then directly translate into an improved data efficiency, i.e., the ability to achieve the same level of performance as supervised training while using only 50-60\% of the labeled data. The original CPC model, as well as the one used in \citep{Mehari:2021Self}, relied on a LSTM model \citep{hochreiter1997long} as predictive model to perform forecasting in latent space, which again poses the question in how far more powerful predictive models, such as causal SSMs, can further improve these results.

\section{Methods}

\subsection{Models}
\noindent\textbf{Structured State Space Models}
Structured State Space Models (SSMs) were introduced in \citep{gu2021efficiently} showing outstanding results on problems that require capturing long-range dependencies. The model consists of stacked \textit{S4 layers} that in turn draw on state-space models, frequently used in control theory, of the form
\begin{equation}
\begin{split}
	x'(t)=&A x(t) + B u(t)\,, \\
	y(t)=&C x(t) + D u(t)\,,
\end{split}
\label{eq:state_space}
\end{equation}
that map a one-dimensional input $u(t)\in\mathbb{R}$ to a one-dimensional output $y(t)\in \mathbb{R}$ mediated through a hidden state $x(t)\in {\mathbb{R}}^N$ parametrized through matrices $A,B,C,D$. These continuous-time parameters can be mapped to discrete-time parameters $\bar{A}, \bar{B}, \bar{C}$ for a given step size $\Delta$.
These allow to form the \textit{SSM convolutional kernel},

\begin{equation}
	 \bar{K}(\bar A, \bar B, \bar C) = (\bar C \bar B, \bar C \bar A \bar B, \bar C \bar A^2 \bar B, ...,  \bar C \bar A^{l-1} \bar B)\,,
\end{equation}

that allows to calculate the output $y$ by a  convolution operation, $y = \bar{K} \ast u + D u$. One of the main contribution of \citep{gu2021efficiently} lies in providing a stable and efficient way to evaluate the kernel $\bar{K}$. Second, building on earlier work \citep{NEURIPS2020_102f0bb6}, they identify a particular way, according to HiPPO theory \citep{NEURIPS2020_102f0bb6}, of initializing the matrix $A\in {\mathbb{R}}^{n\times n}$ as key to capture long-range interactions. $H$ copies of such layers parametrizing a mapping from $\mathbb{R}\to\mathbb{R}$ are now concatenated and fused through a point-wise linear operation to form a S4 layer mapping from ${\mathbb{R}}^H \to {\mathbb{R}}^H$, in close analogy to the architecture of a transformer block, where the SSM convolution serves as a replacement for multi-head self-attention. These H copies can be thought of H convolutional filters, which are sequence-to-sequence mappings, parametrized through the state-space \Cref{eq:state_space}.

\noindent \textbf{Supervised model} The model used for supervised training follows the original S4 architecture \citep{gu2021efficiently} and consists of a convolutional layer as input encoder, followed by four \textit{S4 blocks} which are connected through residual connections interleaved with normalization layers, with a global pooling layer and a linear classifier on top. The S4 blocks comprise the S4 layer accompanied by dropout and GeLU activations and a linear layer.

S4-Layers are designed to model long-term time series data by capturing the dependencies and patterns in the sequential data. They are based on the state-space formulation, see \Cref{eq:state_space}, which describes the evolution of hidden states over time. As mentioned before, the output of the S4 layer can be calculated by a simple convolution with the SSM convolutional kernel, which has been shown to be equal to iteratively solving the equation to update the hidden states through time. This allows the model to capture the sequential patterns and relationships within the ECG signals using a highly parallelizable operation. Hence, by leveraging the state-space (equation \ref{eq:state_space}), S4-Layers enable the modeling of long-term dependencies and complex temporal dynamics in the ECG data. We refer to \citep{gu2021efficiently} for details.

\begin{figure}
	\centering
	\includegraphics[width=\columnwidth]{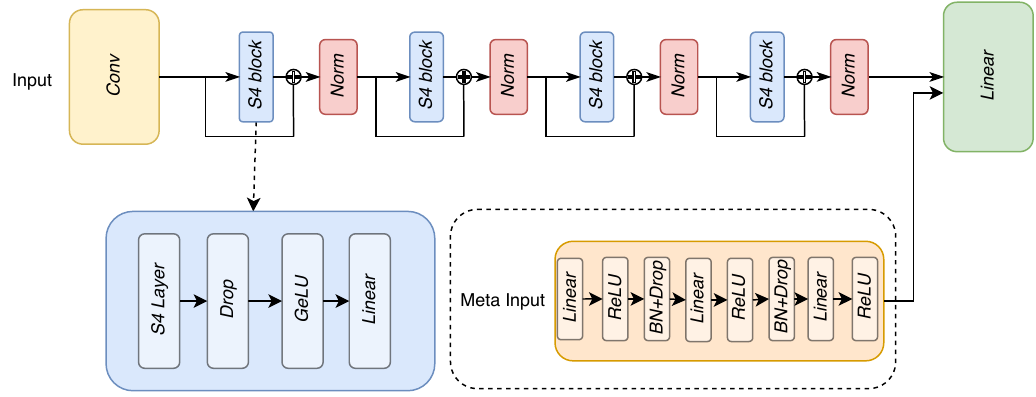}
	\caption{The model used during supervised training follows the original S4 architecture \citep{gu2021efficiently}. The model consists of a convolutional layer as input encoder, followed by four S4 blocks which are connected through residual connections with a normalization layer. The prediction is obtained from a linear layer following a mean pooling layer. In the setting where we also use the patient-specific metadata (of PTB-XL), we include it in the model through \textit{meta head} (depicted in the dotted box at the bottom) that receives the metadata as input. Its output is concatenated with the pooled features representation of the signal itself and passed as input to the final classification layer.
	}
	\label{fig:model}
\end{figure}

The architecture is summarized schematically in Figure~\ref{fig:model}. We refer to this model as the \textit{S4 model} We distinguish causal/masked and bidirectional variants to demonstrate the impact of bidirectional context.

In this work, we also aim to quantitatively explore the impact of including patient-specific metadata on the prediction. To this end, we make use of age, sex, height and weight provided as metadata as part of the PTB-XL dataset. We impute missing values in the sex, height and weight columns using median values inferred from the training set. In all three cases, we include additional binary columns to indicate whether an imputation was applied in the respective columns. We process a total of seven static input features through a simple three-layer multi-layer perceptron (MLP) with ReLU activations and 64 hidden units per layer, which is a negligible number of parameters in comparison to the typically parameter-heavy feature extractors for the raw ECG signals. To improve generalization, we interleave these layers with batch normalization and dropout layers. We concatenate the output of this three-layer neural network with the pooled signal representation extracted from the respective signal classifiers considered in isolation before and pass the latter to the final classification layer to obtain the final classification output. Combining the information from signal and metadata only before the final classification layer is an example of ``intermediate fusion'' approach in the multimodal learning literature \cite{8103116}. We leave the exploration of other fusion schemes to future work and focus on demonstrating that the inclusion of patient metadata leads to consistent performance improvements compared to the synthetic benchmarking case of a classifier operation on raw signals alone.

\noindent \textbf{Self-supervised model} The idea of the contrastive predictive framework \citep{oord2018representation} is to learn informative representations for downstream tasks by solving a forecasting task in latent space. Here, a (causal) prediction model is supposed to predict a latent representation a few time steps ahead from the latent representations observed thus far. We mostly follow the self-supervised learning setup described in  \citep{Mehari:2021Self}. We briefly describe its most important aspects: The signal is processed by a series of four convolutional layers with kernel size 1 and 512 filters. This is an important step since the CPC pretraining is set as a forecasting task in latent space. Unlike in the speech domain, there is no necessity to reduce the temporal resolution, as the input sampling frequency of 100Hz is already considerably lower than typical sampling frequencies of 16kHz in the speech domain. The so defined latent representations serve as targets for the forecasting task in the CPC framework. The task is to minimize a contrastive objective, the InfoNCE loss \citep{oord2018representation},
\begin{equation}
	\label{eq:infonce}
	\mathcal{L}_\text{NCE} = -\mathbbm{E}_X\left[\log \frac{f_k(x_{t+k},c_t)}{\sum_{x_j\in X} f_k(x_j,c_t)}\right]\,,
\end{equation}
where $f_k(x_{t+k},c_t)=\exp(z_{t+k}\dot MLP(c_t))$ models the conditional probability $p(x_{t+k}|c_t)$ between encoded feature representation $g(x_{t+k})\equiv z_{t+k}$ and the forecast implemented as 2-layer MLP operating on the output $c_t$ of a causal predictor model summarizing the sequence up to time step $t$. Here, $g$ refers to the encoder, $x_t$ to the input at time step $t$, $z_t$ to the corresponding latent representation and $X$ to the set of random samples drawn from $p(x_{t+k}$). In our case, we use the output of the SSM predictor as $c_t$. In practice, the negative samples for the evaluation of the denominator in Equation~(\ref{eq:infonce}) are drawn from a single mini-batch. In this particular case, we even draw the negatives from the same sequence as the positive sample. Due to the nature of the contrastive forecasting task, the model architecture has to be causal, i.e.\, the predictor has to be an autoregressive model that processed in a unidirectional fashion. This generally results in slight performance losses compared to the corresponding bidirectional model in a supervised context, which, however, typically gets overcompensated through the self-supervised pretraining step.

\begin{figure}
	\centering
	\includegraphics[width=\columnwidth]{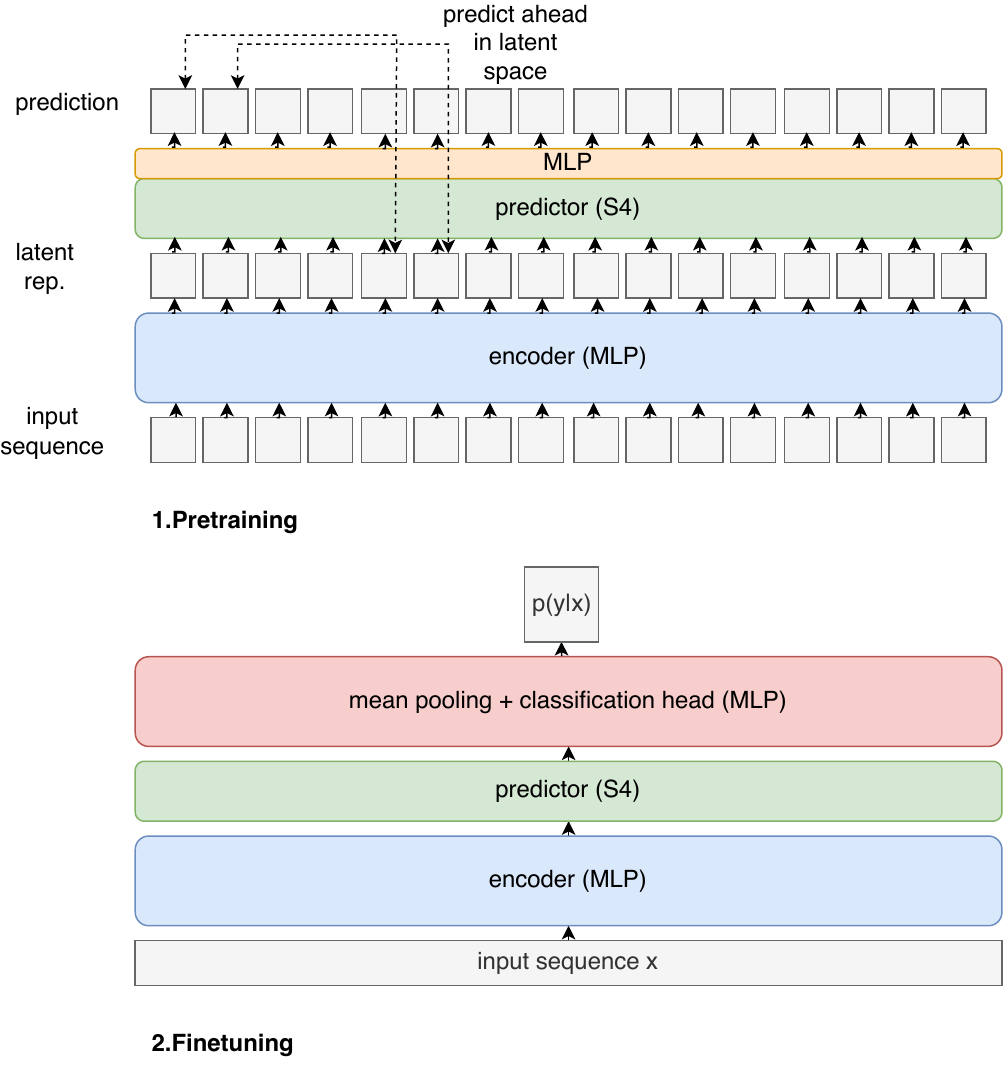}
	\caption{Schematic representation of the pretraining and finetuning procedures followed in this work.
	}
	\label{fig:schematic}
\end{figure}

The most important component is the mentioned predictor model that aggregates the latent representations seen up to a certain point in time to produce a forecast for the latent representation, in this case 12 steps i.e.\ 120ms into the future. Unlike in \citep{Mehari:2021Self}, where the authors used a two-layer LSTM model, we use four S4 blocks in this work. The final prediction is provided after processing through two fully connected layers. During finetuning, the latter are discarded and replaced by mean pooling and a linear classification head. It is worth mentioning that this constitutes another difference to \citep{Mehari:2021Self}, as they use concat pooling and a non-linear classification head. However, we follow the training methodology put forward in \citep{Mehari:2021Self} for the finetuning and first freeze all weights except those in the classification head and subsequently finetune the whole model, except that we refrain from using discriminative learning rates. A schematic overview of the architecture used during pretraining and finetuning is presented in \Cref{fig:schematic}. In the context of self-supervised learned models, we use the notation \textit{LSTM with FCE} to refer to the model from \citep{Mehari:2021Self} and \textit{S4 with FCE} to the architecture proposed in this work, where \textit{FCE} stands for fully-connected encoder and refers to the use of four convolutional layers with kernel size 1 as opposed to a single convolutional encoder layer in the original S4 model.

\subsection{Datasets and experimental procedure} 
To evaluate the effectiveness of our proposed method, we conducted tests using two large 12-lead ECG datasets that are publicly available. These datasets are known as PTB-XL\citep{Wagner:2020PTBXL,Wagner2020:ptbxlphysionet,Goldberger2020:physionet} and Chapman\citep{Zheng2020}. In the case of PTB-XL, we followed a previously established benchmarking methodology\citep{strodthoff2020deep}, where the first eight label-balanced folds of the dataset are used for training, the ninth for validation and the tenth for testing in a multi-label classification task. We assessed the performance on a label set consisting of 71 different labels that cover a wide range of diagnostic, form-related, and rhythm-related statements. As performance metric, we use the macro AUC on the test fold. For the Chapman dataset, we focused on the primary annotations that are related to rhythm statements. To ensure a fair comparison with the PTB-XL dataset, we also divided the Chapman dataset into 10 label-balanced folds (8 for training, 1 for validation, and 1 for testing) and also used the macro AUC as the performance metric. Moreover, we filtered out statements that occurred fewer than ten times to ensure that each fold had at least one sample of each ECG statement, resulting in a reduced set of 48 statements from the original 68. It's important to note that both the PTB-XL and Chapman datasets are widely used and publicly available, representing a diverse range of patient cohorts and various pathologies. Therefore, they are suitable for thoroughly evaluating the effectiveness of ECG classification algorithms.\\
For a thorough assessment of self-supervised learning, we leverage large collections of publicly available ECG datasets. The first one, \textit{All2020}, includes the training data from the Computing in Cardiology Challenge 2020 \citep{PerezAlday2020} in a addition to the \textit{Chapman} dataset and the \textit{CODE} test set \citep{Ribeiro2020}. This dataset ensures the comparability to \citep{Mehari:2021Self}, where the same dataset was used for pretraining. Going beyond this, we compiled an even larger pretraining dataset, \textit{All2021}, which includes the Computing in Cardiology Challenge 2021 \citep{reyna2021will} training set as well as the \textit{CODE} test set and comprises in total 89080 samples. It is worth noting that the challenge datasets are by themselves collected from various sources. The \textit{PTB-XL}  dataset is contained both in the 2020 and the 2021 version and the \textit{Chapman} dataset is part of the 2021 dataset. We summarize all datasets used in this study again in \Cref{tab:datasets}.\\
For the experiments involving S4 layers, we use a batch size of 32, $N=8$ for the state dimension in the S4 layers (as optimal value identified on the \textit{PTB-XL}  validation set) and train the model with a constant learning rate schedule and a learning rate $lr=0.001$ for 50 epochs with the AdamW Optimizer \citep{adamw2017}. The supervised training task represents a multi-label classification task and we use binary crossentropy as objective function. If not stated otherwise, we train models on input sequences of length 2.5s in order to remain comparability to results in the literature. During training, subsequences are randomly cropped from the input record. During test time, we use test-time-augmentation: 
We create ten equidistantly overlapping crops from the original sample, of the same length as the input size used in training, such that the entire sample is covered. The average of the ten respective output probabilities is used as the final prediction for the entire sample. 

\begin{table}
	\centering
	\caption{Overview over the datasets used in this study.}
	\begin{tabular}{ll}
		\toprule
		\bfseries dataset  & \bfseries \# samples  \\
		\midrule
		\multicolumn{2}{c}{Evaluation/Supervised training}\\\midrule
		\textit{PTB-XL} \citep{Wagner:2020PTBXL} & 21,837 \\
		\textit{Chapman} \citep{Zheng2020}  & 10,646\\\midrule
		 \multicolumn{2}{c}{Self-supervised pretraining}\\\midrule
		PTB-XL \citep{Wagner:2020PTBXL} & 21,837 \\
		 \textit{All2020}=\textit{CinC2020}+\textit{Chapman}+\textit{CODE}-Test  & 54,574=43,101+10,646+827 \\
		 \textit{All2021}=\textit{CinC2021}+\textit{CODE}-Test & 89080=88,253+827 \\
		\bottomrule
	\end{tabular}
	\label{tab:datasets}
\end{table}\

%% file: jbhi_sections/results.tex
\section{Results}
\noindent \textbf{SSMs outperform the current supervised state-of-the-art}
As mentioned above, state-of-the-art approaches in deep-learning-based ECG analysis mainly rely on modern convolutional architectures. As a representative example, we use a model with \textit{xresnet1d50} architecture that was shown to lead to competitive results compared to the state-of-the-art at that time \citep{Mehari:2021Self}. We also compare it to a recurrent neural network with a fully-connected feature extractor (\textit{LSTM with FCE}) \citep{Mehari:2021Self} that showed the best reported supervised performance on \textit{PTB-XL} to date. In Table~\ref{tab:supervised}, we compare these models to the \textit{S4 model} based on comprehensive ECG classification tasks on the \textit{PTB-XL} and \textit{Chapman} datasets. Here and in the following, we report mean and standard deviation of the test set scores over 10 runs using a concise error notation where e.g.\ 0.9175(39) signifies $0.9175\pm 0.0039$. On \textit{PTB-XL}, the  \textit{S4 model} surpasses all baseline methods in a statistical significant manner, see below for a detailed description. The fact that PTB-XL has developed into a reference dataset for the benchmarking of ECG classification models, allows to compare the proposed method also to a variety of other approaches on the dataset. Most notably, the \textit{S4 model} outperforms recently proposed transformer/attention-based methods such as \cite{zhang2021multi,li2022dual}. However, the results also highlight the necessity of measuring and reporting statistical uncertainties of the results. Only those allow to assess the significance of the reported results. Interestingly, the ranking of the algorithms is largely consistent on the \textit{Chapman} dataset but the differences between the different approaches are not significant in this case. This might be explained by the fact that all models achieve very high predictive performance on this dataset with macro AUC values beyond 0.98, which is not the ideal situation if one aims to quantify performance differences in a statistically significant manner.\\
\noindent \added{\textbf{Statistical significance}}
As a final comment on the significance of the improvements achieved, it should be noted that (a) ECG statement prediction on PTB-XL is so far the only widely accepted benchmarking setting for comprehensive ECG prediction, and (b) improvements in a target metric are difficult to assess solely on the basis of their numerical value, but should rather be judged on the significance of the improvements over the previous state-of-the-art.
To this end, we have to refine our notion of statistical and systematical uncertainty measures. Following the methodology used in \citep{Mehari:2021Self}, we consider two sources of uncertainty, the statistical uncertainty due to the randomness of the training process, which can be assessed through multiple training runs, and the uncertainty due to the finiteness and the specificity of the label distribution of the test set. We address the latter via empirical bootstrapping ($n_\text{iter}=1000$ iterations) on the test set. Comparing two particular trained models, we consider the performance difference to be statistically significant if the bootstrap 95\% confidence intervals for the performance difference does not overlap with zero. We address the uncertainty due to the stochasticity of the training process by training $n_\text{runs}=10$ for each of the models we aim to compare. For each of the $n_\text{runs}^2$ comparisons, we assess the statistical significance via bootstrapping as defined above. Finally, we define a model to perform statistically significantly better/worse in case that 60\% of the model comparisons turn out to be statistically significantly better/worse. Like the significance level, this threshold can be chosen at will as long as it exceeds 40\% for consistency reasons \citep{Springenberg2022Histo} and is related to the amount of uncertainty  one is willing to tolerate  due to fluctuations across training runs.

% reporting means:
\begin{table}[ht]
	\centering
	
	\caption{Comparing supervised performance of the state-of-the-art models on two large ECG datasets.  The asterisk stands for statistically significant better performance than \textit{LSTM with FCE} (as the previous state-of-the-art with quantified statistical uncertainty). Results marked with ${}^\dagger$ were obtained from own experiments, all remaining values were taken from the literature, for which unfortunately no statistics over multiple runs is available.}
	\label{tab:supervised}
	\begin{tabular}{lll}
		\toprule
		&\multicolumn{2}{c}{dataset}\\
		& \textit{PTB-XL} &  \textit{Chapman}\\
		
		\midrule 
		\textit{ViT} \citep{li2021bat} & 0.862 &  - \\
		\textit{BaT} \citep{li2021bat} & 0.905 &  - \\
		\textit{inception1d} \citep{strodthoff2020deep} & 0.925 &  - \\
		\textit{xresnet1d101} \citep{strodthoff2020deep} & 0.925 &  - \\
		\textit{ensemble} \citep{strodthoff2020deep} & 0.929 &  - \\
		\textit{multi-period attention} \cite{zhang2021multi} & 0.932 & - \\
		\textit{DLTB-ECG} (signal only) \citep{li2022dual} & 0.934 &  - \\
		\textit{xresnet1d50}${}^\dagger$ \citep{strodthoff2020deep} & 0.9286(28) &  0.9805(39) \\
		\textit{LSTM with FCE}${}^\dagger$ (causal) \citep{Mehari:2021Self} &0.9295(31) &0.9854(12) \\%\hline
		\textit{S4 model}${}^\dagger$ (this work)& \textbf{0.9417(16)*} &  \textbf{0.9876(11)}\\
		
		\bottomrule	
		
	\end{tabular}
	
\end{table}

\begin{figure*}
	\centering
	\includegraphics[width=.8\textwidth]{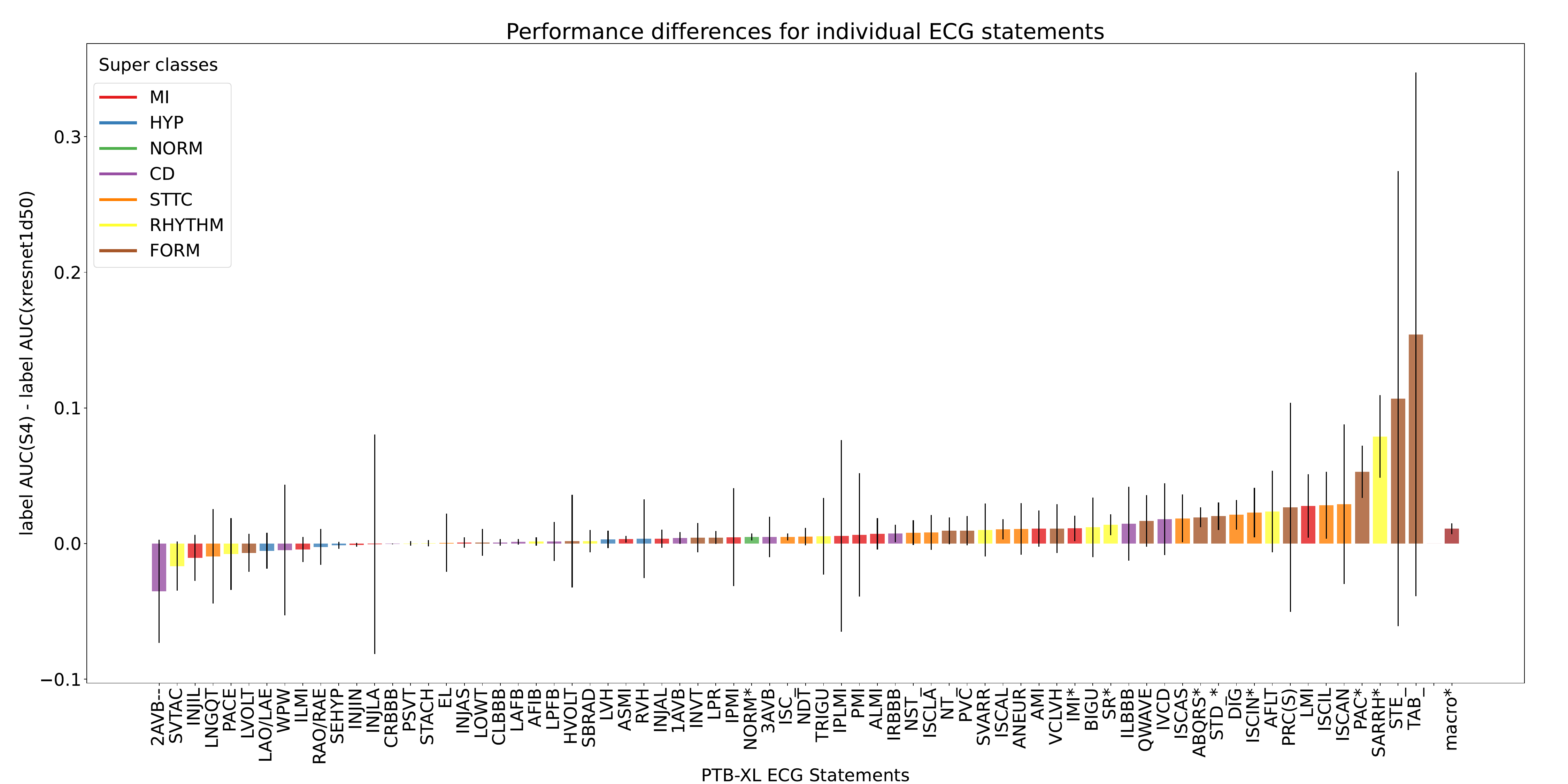}
	\caption{\replaced{Comprehensive comparison of the performance difference, per pathology, between the \textit{S4} and \textit{xresnet1d50} models at a sampling rate of 100Hz. The x-axis enumerates the ECG statements (as well as macro AUC across all statements) in the PTB-XL dataset, see \cite{Wagner:2020PTBXL} for details, while the y-axis indicates the difference in AUC values. The bars, color-coded according to the respective super classes, indicate the median performance difference over 100 comparisons comparing trained 10 S4 models and 10 xresnet1d50 models. The thin vertical lines represent the respective interquartile ranges, as a measure of how much the differences vary. Additionally we assess the statistical significance of the performance improvements through bootstrapping on the test set, see the main text for details, where an asterisk(hyphen) next to an ECG statement indicates that the \textit{S4} model performs statistically significantly better(worse) than the \textit{xresnet1d50} model.}{Comparison of the performance of the \textit{S4} and \textit{xresnet1d50} models  at a sampling rate of 100Hz. We plot label AUC(S4 model)-label AUC(xresnet1d50). The asterisk means that the \textit{S4 model} performs statistically significantly better and -- that it performs statistically significantly worse than the \textit{xresnet1d50} model on the respective label.}}
	\label{fig:stat_diffs}
	
\end{figure*}
\label{app:paths}
In Figure~ \ref{fig:stat_diffs}, we show a bootstrap comparison between the \textit{S4 model} architecture and the \textit{xresnet1d50} architecture on the level of individual ECG statements. The colored bars represent the median values and the black bars the standard deviation of the $n_\text{runs}^2$ medians of the $n_\text{iter}$ AUC difference comparisons per model combination. Labels on which the \textit{S4 model} architecture performs statistically better(worse) are marked by  *(--). The plot reveals that despite high median values for the difference of some label AUCs, like e.g for non-specific ST Elevation (STE) or T-wave abnormality (TAB), the \textit{S4 model} architecture does not necessarily perform statistically better on those labels, as the difference varies strongly over combinations of different runs. On the other hand though, there are pathologies, where the median AUC difference is close to zero but with such a low variance, that these differences are statistically significant, as it is the case for e.g.\ healthy ECG signals (NORM) or inferior myocardial infarctions (IMI). Summarizing, we find statistical significant improvements for 8 (SARRH, PAC, ISCIN, STD\_, ABQRS, SR, IMI, NORM) of the 71 ECG statements, while only 2AVB showed a consistent decrease in performance.  

\noindent \textbf{SSMs allow inference at unseen sampling rates}
A compelling aspect of state space models is that, due to the continuous character of the state-transition matrix $\textbf{\textit{A}}$ in Equation~(\ref{eq:state_space}), the model can be evaluated on data that was sampled at a different rate than the training data,  simply by adjusting the step size in the discretization step during test time. Table \ref{tab:model_selection} depicts a cross-evaluation matrix, in which we trained and cross-evaluated the \textit{S4 model} on 100, 200 and 500Hz. We see no or just minor losses in performance when varying test from train sampling rates, even if the sampling rates differ by a factor of 5. This is a particular asset of SSM models as it avoids the necessity to resample the data, which might otherwise be a source of additional systematic uncertainties. 

% with mean and std:
\begin{table}[ht]
	\centering
	\caption{
		Comparing supervised performance of the \textit{S4 model} trained/tested on different sampling rates.}
	\label{tab:model_selection}
	\begin{tabular}{lllll}
		\toprule
		&&\multicolumn{3}{c}{test}\\
		&& 100Hz & 200Hz & 500Hz\\
		
		\midrule
		
		\multirow{3}{*}{\rotatebox[origin=c]{90}{train}}
		&100Hz & 0.9417(35) & 0.9418(36) & 0.9416(36)\\
		
		&200Hz    & 0.9414(16)  & 0.9416(16)& 0.9417(16)\\
		
		&500Hz   & 0.9415(18) & 0.9421(19)& 0.9421(19) \\
		\bottomrule	
	\end{tabular}
	
\end{table}

% reporting means:
\begin{table}[]
	\centering
	\caption{Comparing downstream performance (macro AUC on the PTB-XL/Chapman test set for the most finegrained level of the label hierarchy) after self-supervised pretraining on different datasets. The asterisk stands for statistically significant better performance than its supervised counterpart. The first result was taken from \citep{Mehari:2021Self}. FCE stands for fully-connected encoder.}
	\begin{tabular}{llll}
		\toprule
		\bfseries Pretraining & \bfseries model  & \bfseries \textit{PTB-XL} & \bfseries  \textit{Chapman}\\
		\midrule
		%PTB-XL &  4FC+2LSTM+2FC \citep{Mehari:2021Self}& 0.9395(11)&\\
		\textit{All2020}  & \textit{LSTM with FCE}& 0.9418(14)*&  0.9887(07)\\\hline
		%PTB-XL  & 4FC+4S4+1FC (this work) & 0.9436(12)&\\
		\textit{All2020}  & \textit{S4 with FCE} (this work)&  0.9436(16)*& 0.9887(06)\\
		\textit{All2021} & \textit{S4 with FCE} (this work) & \bfseries 0.9445(19)*& \bfseries 0.9892(09)\\
		\bottomrule
	\end{tabular}
	\label{tab:self}
\end{table}

\begin{figure}[ht]
	\centering
	\includegraphics[width=\columnwidth]{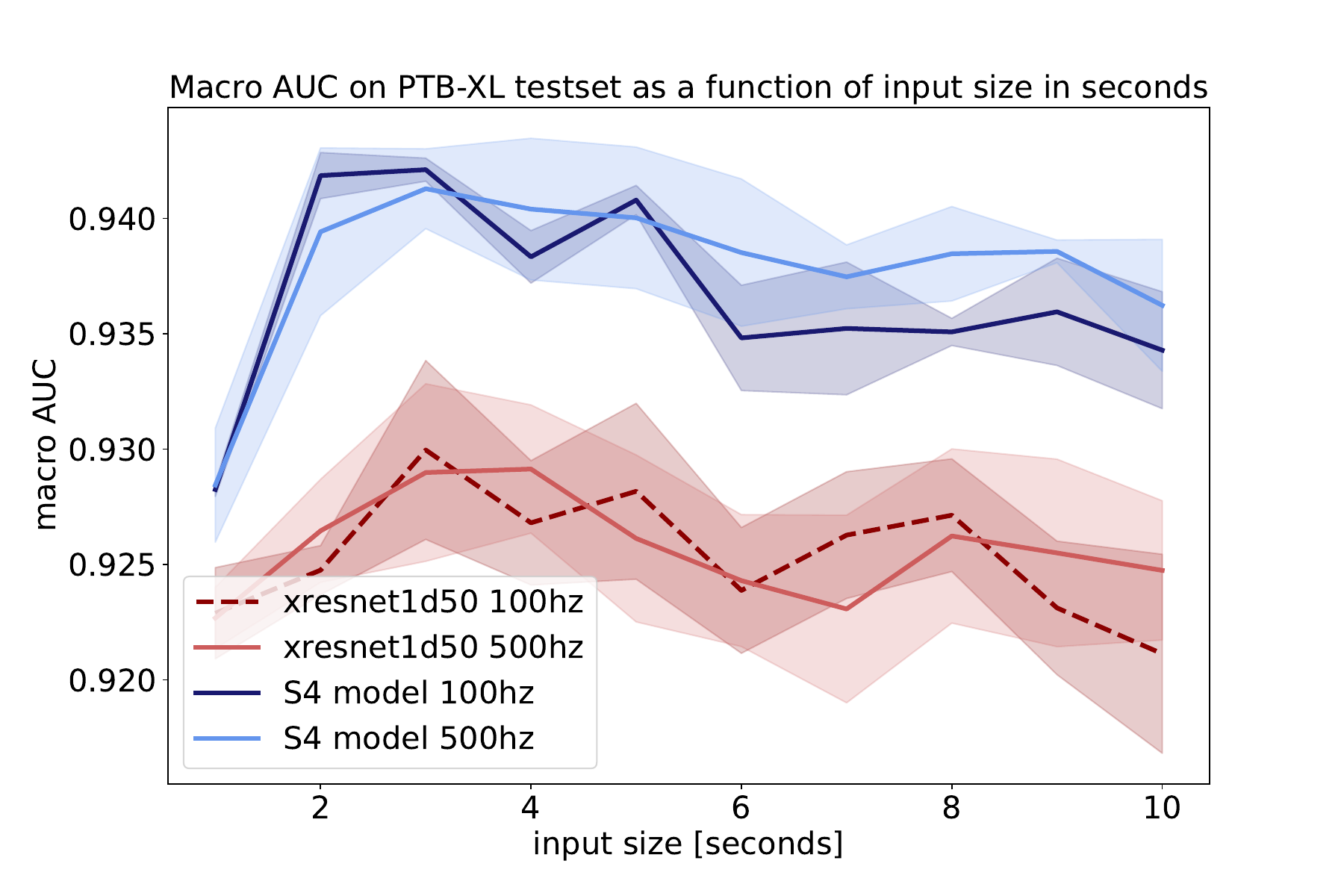}
	\caption{Model performance in dependence of (physical) input size for two convolutional models (\textit{xresnet1d50}) at 100/500Hz as compared to two SSMs (\text{S4 model}) at 100/500Hz.}
	\label{fig:input_size}
\end{figure}

\noindent \textbf{No long-term dependencies in ECG data beyond 3s}
In this paragraph, we aim to clarify in a quantitative way, how long-ranged interactions are actually present in ECG data, which is a long-standing question that has not been systematically addressed so far. We address this question in terms of the size of the input window that is passed to the model (while performing test-time-augmentation, i.e.\, combining information from different segments for the sample-level prediction at all times). We believe that this question has not be answered in an unbiased way so far due to the inability of prior architectures to capture long-term dependencies in the data without adjusting hyperparameters such as kernel sizes etc.\ in the case of convolutional models.
In Figure~\ref{fig:input_size}, we investigate the model performance (macro AUC measured on the \textit{PTB-XL} test set) as a function of input size for two convolutional models using input data sampled at 100Hz and 500Hz and two SSMs using the same kind of input data. As described above, we train models on various input sizes, measured in physical units for comparability, and obtain aggregated predictions for the full samples during test time by taking the average of ten input windows that are consequently moved through the signal with varying stride. As a first observation from Figure~\ref{fig:input_size}, we see that the performance gap between convolutional models and structured state space models, which was already apparent at an input size of 2.5s in Table~\ref{tab:supervised}, persists across all input sizes. Second, within each model architecture, the results from input data sampled at 100Hz as compared to 500Hz largely overlap. This already puts into question the potential advantage of using input data at 500Hz for ECG classification purposes. We will revisit this question in a more detail in the next paragraph. Third, the plot shows an interesting dependence on the input size that is qualitatively consistent across model architectures: The performance from aggregated model predictions shows a peak around input sizes around 2-3s. This hints at the fact that for ECG classification tasks based on short 12-lead ECG data, the ability of the model to explicitly capture long-range interactions beyond about 3s is not beneficial. On the contrary, the models seem\replaced{s} to profit more from averaging overlapping predictions from different sliding windows. This observation is very much in line with the fact that most pathologies affect all beats equally (with a few exceptions such as premature ventricular contractions) and the fact that for average heart rates between 60 and 100bpm, a sliding window of 3s already contains 3-5 beats. This question is obviously completely independent from the question of capturing long-term dependencies in long-term ECGs with rhythm changes within the sample and should be revisited in future work.

\noindent\textbf{No significant advantages from sampling frequencies beyond 100Hz}
We also revisit the question of the comparison between sampling frequencies at 100Hz vs. 500Hz in a more statistically rigorous manner based on the methodology presented above. At a input size of 2.5s and for fixed model architecture, we find no statistically significant performance difference between both sampling frequencies. This applies equally well to the level of individual label AUCs, in an analysis analogous to the one carried out for the comparison between the S4 and xresnet model  above. We want to stress that this statement obviously depends strongly  on the label distribution of the dataset under consideration, in the sense that there might be systematic improvements for certain ECG statements that do not turn out to be statistically significant due to large fluctuations as a consequence of small sample sizes.

\noindent\textbf{CPC with SSMs predictor outperform the current self-supervised state-of-the-art}
As a final experiment, we also study the impact of SSMs in the context of self-supervised pretraining. The results compiled in Table~\ref{tab:self} reveal a statistically significant improvement in terms of downstream performance compared to the previously best result (\textit{LSTM with FCE}) for the identical pretraining dataset \textit{All2020}. Enlarging the pretraining dataset even further, as for \textit{All2021}, leads to a further performance improvement, reaching 0.9445(19), the highest performance reached so far on \textit{PTB-XL}. In all cases, we find statistically significant improvements on \textit{PTB-XL} compared to the corresponding model trained in a supervised fashion. To show that the achieved results also generalize to other datasets, we repeat the same downstream analysis using the same pretrained model for the  \textit{Chapman} dataset and find qualitatively similar results. At this point, it is hard to assess if the improvements in downstream performance are primarily consequences of the improved performance of the model architecture, as seen in supervised training. In any case, replacing LSTMs by S4 layers in architectures used for self-supervised pretraining represents a promising step also for other data modalities such as speech.

\noindent\textbf{Incorporation of patient-specific metadata enhances predictive performance}
In \Cref{tab:meta}, we report the performance evaluation of the incorporation of patient-specific metadata. It reveals a consistent improvement compared to the performance of the corresponding model operating on input signals only across all models. The convolutional \textit{xresnet1d50} model benefits most  from the inclusion of metadata. The pretrained SSM model \textit{S4 with FCE} (pretrained on \textit{All2021}) reaches on overall performance high at a macro AUC of 0.9463(08). We urge the community to also adopt this clinically more relevant task as a benchmark. To be able to assess the resilience of the results, we strongly urge to report also statistical fluctuations over multiple training runs.

% reporting mean
\begin{table}[ht]
	\centering
	
	\caption{Comparing supervised performance of the state-of-the-art models on PTB-XL that incorporate patient-specific metadata as compared to models operating on signals only. The notation follows \Cref{tab:supervised}.}
	\label{tab:meta}
	\begin{tabular}{lll}
		\toprule
		&\multicolumn{2}{c}{\textit{PTB-XL}}\\
		& signal & signal \& meta\\
		
		\midrule 
		\textit{xresnet1d50}${}^\dagger$ (this work) & 0.9286(28)& 0.9388(26)\\
		\textit{DLTB-ECG} \citep{li2022dual} & 0.934 &0.942  \\
		\textit{S4 model}${}^\dagger$ (supervised, this work)& 0.9417(16)& 0.9434(33)\\
		\textit{S4 with FCE}${}^\dagger$ (pretrained, this work)& \textbf{0.9445(19)}& \textbf{0.9463(08)} \\
		
		\bottomrule	
		
	\end{tabular}
	
\end{table}

\begin{figure}
	\centering
	\includegraphics[width=\columnwidth]{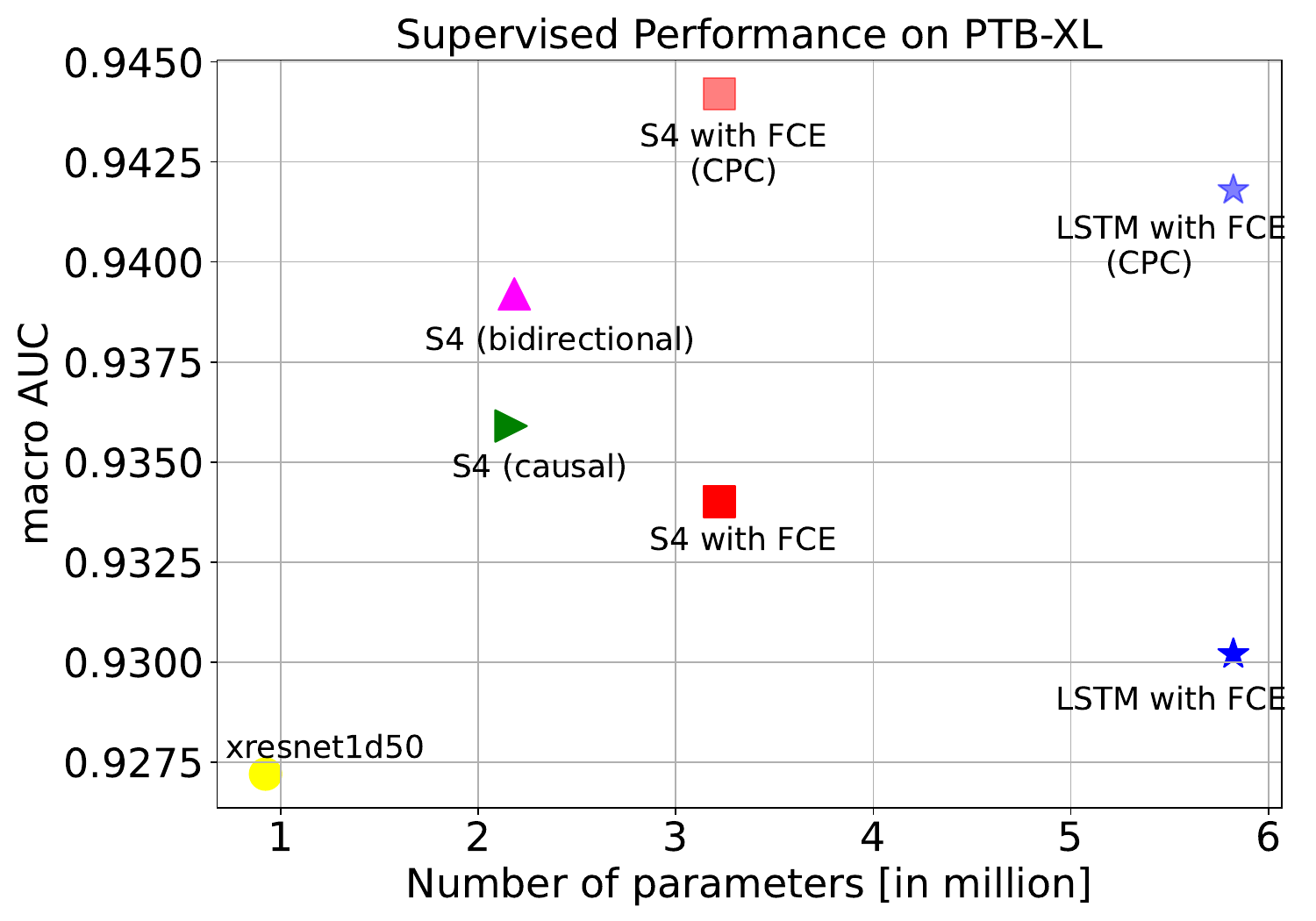}
	\caption{Visual summary: Supervised performance on \textit{PTB-XL} (with and without self-supervised pretraining). Models involving SSMs clearly advance the state-of-the-art while maintaining a smaller parameter budget as previous models using LSTMs as predictors.\textit{FCE} stands for fully-connected encoder and \textit{(CPC)} indicates that the model received self-supervised pretrained using contrastive predictive coding.}
	\label{fig:summary}
\end{figure}

\noindent \textbf{Visual summary} In Figure \ref{fig:performance_contributions} and \ref{fig:summary}, we present visual summaries of the outcome of this study. They show the improvements achieved through the use of structured state space model layers both in the supervised and in the self-supervised domain. In Figure \ref{fig:summary}, we use the number of model parameters to give an impression of the model complexity, although other parameters such as inference time should be considered for a more complete picture  \citep{Deghani2021efficiency}.

